\documentclass[prb,twocolumn, showpacs,preprintnumbers,superscriptaddress]{revtex4}
\usepackage{graphicx}
\usepackage{dcolumn}
\usepackage{bm}
\usepackage{subfigure}
\usepackage[dvips]{color}

\begin{document}

\title{Optical evidence for intermolecular coupling in mixed films of pentacene and perfluoropentacene} 

\author{K. Broch}\affiliation{Universit\"at T\"ubingen, Insitut f\"ur Angewandte Physik, Auf der Morgenstelle 10, 72076 T\"ubingen, Germany}
\author{U. Heinemeyer}\affiliation{Universit\"at T\"ubingen, Insitut f\"ur Angewandte Physik, Auf der Morgenstelle 10, 72076 T\"ubingen, Germany}
\author{A. Hinderhofer}\affiliation{Universit\"at T\"ubingen, Insitut f\"ur Angewandte Physik, Auf der Morgenstelle 10, 72076 T\"ubingen, Germany}
\author{F. Anger}\affiliation{Universit\"at T\"ubingen, Insitut f\"ur Angewandte Physik, Auf der Morgenstelle 10, 72076 T\"ubingen, Germany}
\author{R. Scholz}\affiliation{Technische Universit\"at Dresden, Institut f\"ur Angewandte Photophysik, 
01062 Dresden, Germany}
\author{A. Gerlach}\affiliation{Universit\"at T\"ubingen, Insitut f\"ur Angewandte Physik, Auf der Morgenstelle 10, 72076 T\"ubingen, Germany}
\author{F. Schreiber}\email{frank.schreiber@uni-tuebingen.de}\affiliation{Universit\"at T\"ubingen, Insitut f\"ur Angewandte Physik, Auf der Morgenstelle 10, 72076 T\"ubingen, Germany} 

\date{\today}

\begin{abstract}
We present optical absorption spectra of mixed films of pentacene (PEN) and perfluoropentacene (PFP) grown on $\mathrm{SiO_{2}}$. We investigated the influence of intermolecular coupling between PEN and PFP on the optical spectra by analyzing samples with five different mixing ratios of PFP:PEN with variable angle spectroscopic ellipsometry (VASE) and differential reflectance spectroscopy (DRS). The data show how the spectral shape is influenced by changes in the volume ratio of the two components. By comparison with the pure film spectra an attempt is made to distinguish transitions due to intermolecular coupling between PEN and PFP from transitions caused by interactions of PEN (PFP) with other molecules of the same type. We observe a new transition at 1.6~eV which is not found in the pure film spectra and which we assign to coupling of PFP and PEN. 
\end{abstract}


\maketitle 
%
%
%
\section{Introduction}
In recent years heterostructures of organic semiconductors have gained increased attention because of their applications in organic opto-electronics~\cite{Anthony_2008_ACIE_47_452}. In addition to layered heterostructures, also bulk-heterojunctions, i.e. co-deposited mixed films, became important in particular for increasing the donor-acceptor interface in solar cells. In such systems the effects of intermolecular coupling on the optical, electronic and structural properties are fundamental, in particular for applications where the mixture of the molecules serves as a p-n-junction. Those effects can be efficiently studied in systems that exhibit intermixing on a molecular level. Perfluorination is a suitable method for creating n-conducting compounds~\cite{Medina_2007_JCP_126_111101, Chen_2005_CPL_401_539, Delgado_2009_JACS_131_1502} that have the analogous structure as their parent (protonated) compound which offers good chances for efficient intermixing under suitable conditions. Therefore, mixed systems of perfluorinated and protonated molecules have been investigated~\cite{Meyer_2003_ACIE_42No11_1210},
including mixtures of pentacene (PEN) and perfluorinated pentacene (PFP)~\cite{Salzmann_2008_L_24_7294, Duhm_2009_APL_94_33304a, Salzmann_2008_JACS_130_12870, Hinderhofer_submitted} and it was already demonstrated that they form mixed crystal structures on the molecular level~\cite{Meyer_2003_ACIE_42No11_1210, Salzmann_2008_L_24_7294, Hinderhofer_submitted}. The optical behavior of mixtures of PFP and PEN is yet to be analyzed and it is of high interest from a fundamental perspective since interactions between molecules can influence the spectra~\cite{Birks_1970_J.P.B._3_1704}.

In this paper we present optical absorption spectra of thin films containing PFP and PEN at various mixing ratios. The effects of changing the PFP:PEN-ratio are illustrated and a decomposition approach into single-component subbands is tested. We find evidence for strong intermolecular coupling between PEN and PFP that severely influences the absorption spectra of the materials investigated.

%
%
\section{Experimental details}
The samples studied were prepared on substrates cut from two 1~mm thick Si(100) wafers with different oxide thicknesses (thermal oxide $d_{\mathrm{ThOx}}$ = 147~nm and native oxide $d_{\mathrm{NtveOx}}$ = 1.9~nm) as described in Ref.~\onlinecite{Heinemeyer_2008_PRB_78_085210}. We grew thin films at a constant substrate temperature of 30~$^{\circ}$C (RT) and with a final film thickness of 20 to 24~nm. Five different nominal mixing ratios of PFP:PEN (1:4, 1:2, 1:1, 2:1 and 4:1) were chosen. Before film growth the temperatures of the evaporation cells for a defined nominal mixing ratio were determined \textit{in situ} with a water-cooled quartz-crystal microbalance (QCM) which was also used to monitor the film thickness and the overall growth rate of $\sim~1$~$\mathrm{\AA}$/min during deposition. After growth the actual mixing ratio was measured with X-ray photoelectron spectroscopy (XPS) and found to be consistent with the nominal ratios within $\leq 6$\%.

%
%
The optical properties of the films were studied \textit{ex situ} with variable angle spectroscopic ellipsometry (VASE)~\cite{Azzam_1977} and \textit{in situ} with differential reflectance spectroscopy (DRS)~\cite{Forker_2009_PCCP_11_2142, Proehl_2005_PRB_71_165207} following procedures outlined in Refs.~\onlinecite{Heinemeyer_2008_PRB_78_085210},~\onlinecite{Heinemeyer_2008_p.s.s.a_205_927} and~\onlinecite{Heinemeyer_2010_P_104_257401a}. On the Si substrate with $d_{\mathrm{NtveOx}}$ = 1.9~nm the DRS-data were obtained at normal incidence in the energy range from 1.5 to 2.8~eV using a deuterium-tungsten halogen-light source (DH-2000, Mikropack) and a fiber optic spectrometer (USB-2000, OceanOptics).
The DRS-signal is defined as~\cite{Proehl_2005_PRB_71_165207}\newline
\begin{equation}
\mathrm{DRS} = \frac{R(d)-R_0}{R_0}
\label{eq:DRS}
\end{equation}
where $R(d)$ corresponds to the reflectivity of the substrate covered with a film with thickness $d$ and $R_0$ denotes the reflectivity of the bare substrate.
Due to the normal incidence geometry of our measurements only the in-plane component of the dielectric function can be studied with DRS. The data analysis was performed with a Gaussian oscillator model to describe the dielectric function of the material~\cite{Heinemeyer_2010_P_104_257401a}.

\begin{figure}
\includegraphics[width=9cm]{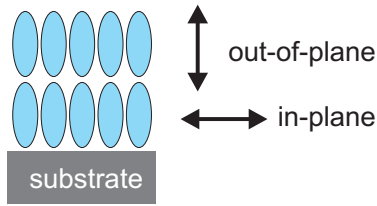}
\caption{(Color online) Definition of the two components of the dielectric tensor describing the optical properties of an uniaxial anisotropic film.}
\label{Fig:e2_Components}
\end{figure}

On both substrates the VASE-data were measured \textit{ex situ} in air with a spectroscopic ellipsometer (M-2000, J.A. Woollam Co.) in the energy range from 1.25 to 3~eV. The angle of incidence relative to the surface normal was varied in steps of 5$^{\circ}$ from 40$^{\circ}$ to 80$^{\circ}$. Performing a multi-sample analysis~\cite{Heinemeyer_2008_p.s.s.a_205_927} we determined the anisotropy of the mixed films with VASE, integrating over an area of $\sim 3$~mm$^{2}$ due to the size of the light spot and therefore averaging over possible inhomogeneities. The mixed films exhibit structural order that leads to uniaxial anisotropy~\cite{Heinemeyer_2008_p.s.s.a_205_927} which is dominated by molecules standing upright~\cite{Salzmann_2008_L_24_7294, Hinderhofer_submitted} similar to the pure films~\cite{Hinderhofer_2007_J.C.P._127_194705}. In case of uniaxial anisotropic films the optical properties are described by two dielectric functions~\cite{Heinemeyer_2008_p.s.s.a_205_927}, i.e. the in-plane and the out-of-plane component of the dielectric tensor, see Fig.~\ref{Fig:e2_Components}.
The comparison of \textit{in situ} DRS- and \textit{ex situ} VASE-measurements revealed slight differences in the in-plane component, probably caused by photo-oxidation~\cite{Kytka_2007_APL_90_131911}, but overall a consistent picture is obtained. 

%
\section{In-plane component of the absorption spectra}

\begin{figure}
\includegraphics[width=9cm]{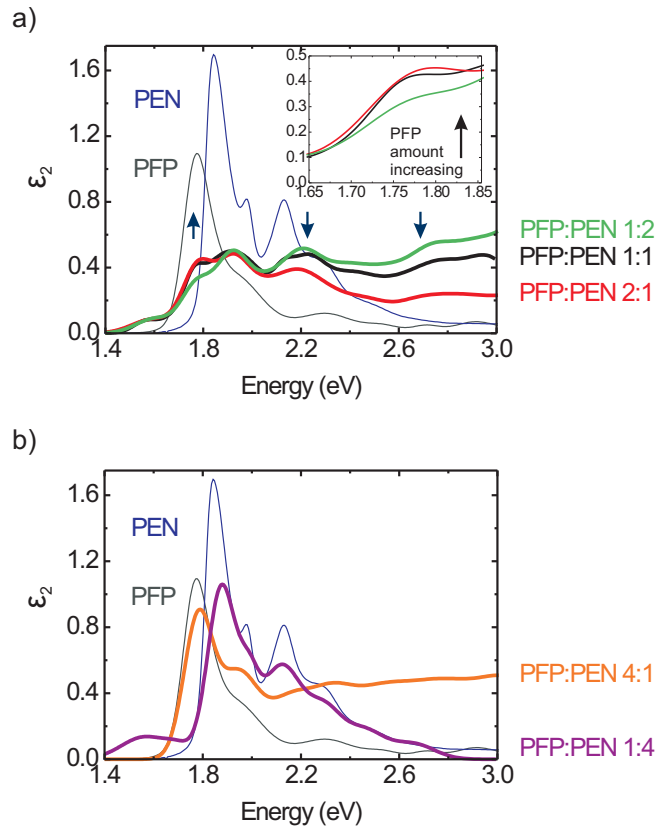}
\caption{(Color online) Imaginary part of the dielectric function (in-plane 
component) of the films with different mixing ratios: 
(a) PFP:PEN 1:1, 2:1 and 1:2 (thick solid lines). 
The arrows are indicating the changes in distinct spectral 
regions with increasing amount of PFP. The inset shows a 
close-up of the spectral region 1.65~eV to 1.85~eV to point 
out the changes within this region. (b) PFP:PEN 4:1, 1:4 (thick solid lines).
For comparison the pure film spectra
of PFP and PEN (thin solid lines) of 
Ref.~\onlinecite{Hinderhofer_2007_J.C.P._127_194705} 
are also shown. The error of the absolute intensities is below 10\%. }
\label{Fig:PFPxPENyinplane}
\end{figure}

%
%
\begin{table*}
\begin{ruledtabular}
\begin{tabular}[b]{c c c c c c}
Mixing ratio & $E_{c}$~[eV] & $E_1$~[eV] & $E_2$~[eV] & $E_3$~[eV] & $E_4$~[eV]\\ \hline
Pure PFP & ~ &~1.75~&~1.94~&~2.28~&~2.48\\ 
PFP:PEN 4:1 &~&~1.78~$\pm$0.01~&~1.94~$\pm$0.01~&~2.15~$\pm$0.03~&~2.31~$\pm$0.09\\ 
PFP:PEN 2:1 &~1.60~$\pm$0.01~&~1.77~$\pm$0.01~&~1.91~$\pm$0.01~&~2.16~$\pm$0.01~&~2.46~$\pm$0.05\\ 
PFP:PEN 1:1 &~1.60~$\pm$0.01~&~1.77~$\pm$0.01~&~1.92~$\pm$0.01~&~(*)~2.20~$\pm$0.03~&~2.41~$\pm$0.09\\
PFP:PEN 1:2 &~1.58~$\pm$0.01~&~1.77~$\pm$0.01~&~1.92~$\pm$0.01~&~2.20~$\pm$0.04~&~2.46~$\pm$0.06\\ 
PFP:PEN 1:4 &~1.60~$\pm$0.02~&~1.88~$\pm$0.01~&~1.98~$\pm$0.01~&~2.12~$\pm$0.01&~2.31~$\pm$0.07\\
Pure PEN & ~ &~1.85~&~1.97~&~2.11~&~2.28\\ 
\end{tabular}
\end{ruledtabular}
\label{Tbl:energypos}
\caption{Energy positions of the most pronounced transitions in the mixed films studied. The energy positions of PFP and PEN are obtained from  Ref.~\onlinecite{Hinderhofer_2007_J.C.P._127_194705}. $E_{c}$ denotes the energy position of the first oscillator observable in the spectra of the mixed films that can tentatively be assigned to charge transfer. $E_1$ describes the energy position of the HOMO-LUMO transition of the pure component film spectra and the energy position of the first strong transition in the mixed film spectra, respectively.
(*) For the 1:1-mixture one can discern two maxima at 2.18 and 2.24~eV; in the table we quote the mean for comparison and consistency.}
\label{Tbl:energypos}
\end{table*}

\begin{figure*}
\includegraphics[width=11cm]{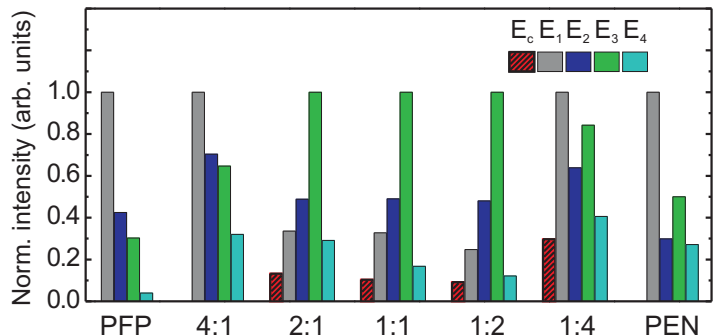}
\caption{(Color online) Relative intensities of the transitions compiled from Tbl.~\ref{Tbl:energypos}, normalized to the transition with the maximum intensity. $E_{c}$, $E_{1}$, $E_{2}$, $E_{3}$ and $E_{4}$ correspond to the respective energy positions noted in the table. For comparison the relative intensities of the pure film spectra are also shown.}
\label{Fig:RelIntensities}
\end{figure*}

Figure~\ref{Fig:PFPxPENyinplane} shows the imaginary part $\epsilon_{2}$ of the in-plane component of the dielectric tensor for the different mixing ratios, obtained with VASE. Since the extinction coefficient $k$ of the material is proportional to $\epsilon_2$, we will refer in the following to the spectra of $\epsilon_{2}$ over photon energy $E$ as the absorption spectra of the films. Here, we find clear evidence for strong intermolecular coupling between PFP and PEN by observing an additional peak at $E_c=1.6$~eV which does not arise from pure PEN or pure PFP films. This new transition is present in all mixtures except the one with mixing ratio PFP:PEN 4:1. A related transition can also be observed in photoluminescence measurements~\cite{Anger_prep}. We tentatively assign this new transition to charge transfer between PFP and PEN similar to the argument put forward in Refs.~\onlinecite{Akaike_2010_AFM_20_715} and~\onlinecite{Clark_2008_JOAP_103_124510} on other mixed systems. In the following we will concentrate on further details of the spectra of the mixed films. For clarity we discuss the mixing ratios close to the equimolar mixture (2:1 and 1:2, see Fig.~\ref{Fig:PFPxPENyinplane}(a)) separately from the other two ratios (4:1 and 1:4, see Fig.~\ref{Fig:PFPxPENyinplane}(b)). From Fig.~\ref{Fig:PFPxPENyinplane}(a) it can be clearly seen that the mixed film spectra are not a linear combination of the single film spectra, which suggests strong intermolecular coupling~\cite{Akaike_2010_AFM_20_715, Clark_2008_JOAP_103_124510}, an assumption further supported by testing also non-linear mixing models, see Sec.~\ref{Chp:NonlinearModels}. The transitions observed are relatively broad compared to the pure film spectra and exhibit no clearly discernible vibronic progression. The shape of the absorption spectra resemble each other, but there are changes observable with increasing amount of PFP, see arrows in Fig.~\ref{Fig:PFPxPENyinplane}(a). The feature at 1.77~eV becomes more pronounced going from a mixing ratio for PFP:PEN of 1:2 to 1:1 and 2:1 (inset in  Fig.~\ref{Fig:PFPxPENyinplane}(a)). As its energy position almost matches the HOMO-LUMO transition of pure PFP, it can be assigned to this compound. On the other hand the feature at 2.2~eV decreases with increasing amount of PFP. Following the same argument and considering that PEN absorbs in this spectral region more strongly than PFP, this absorption band can tentatively be assigned to arise from PEN. 

The shape of the absorption spectra changes strongly when the mixing ratios are close to the single component films (PFP:PEN 4:1 or 1:4, see Fig.~\ref{Fig:PFPxPENyinplane}(b)). In the energy range from 1.7 to 2.1~eV the spectral shape of the absorption spectra of the respective more abundant molecule is clearly dominating, so that the PFP:PEN 4:1 mixture gives a spectral shape essentially resembling pure PFP films and the 1:4-mixture the one of pure PEN. Nevertheless, even here the spectral shape is changed due to inhomogeneities in the film, resulting in a broadening of the peaks which can be deduced from the comparison of the mixed film spectra with the respective single film spectra in Fig.~\ref{Fig:PFPxPENyinplane}(b). Interesting features are the first two pronounced peaks in the spectrum of pure PFP that are assigned to a vibronic progression~\cite{Hinderhofer_2007_J.C.P._127_194705}. The peak at 1.94~eV is more intense in the mixed film spectrum with PFP as abundant molecule compared to the pure film spectrum. 
%
%
%
\subsection{Linear decomposition into single-film subbands}
\label{Subsec:Decomp}
Since there are only few publications concerned with optical spectra of organic blends~\cite{Akaike_2010_AFM_20_715, Clark_2008_JOAP_103_124510, Ruani_1999_SM_103_2392, Datta_2008_TSF_516_7237}, no 'standard procedure' for a data analysis of coupling effects has emerged. As a first approach, we perform a decomposition into single-component subbands. For this purpose, we normalize all spectra to their maximum value and describe the single component spectra by Gaussian oscillators. The shape of the single component spectra is conserved by coupling the oscillators, i.e. keeping the relative energy positions and relative intensities of the oscillators constant. The normalized mixed film spectra are then fitted with the coupled Gaussian oscillators of PFP and PEN. If necessary, additional oscillators are added to describe transitions in the mixed film spectra not resulting from a linear combination of the single component spectra, arising instead from intermolecular coupling. The energy positions of the Gaussian oscillators resulting from the decomposition are compiled in Tbl.~\ref{Tbl:energypos}, revealing the existence of a transition in the mixed films that cannot be described by Gaussian oscillators related to PFP or PEN. For comparison also the energy positions of the transitions of pure films are denoted. In order to analyze the effects of intermolecular coupling on the relative intensities in the spectra of the mixed films we analyzed the in-plane component of the spectra. For this we calculated the intensities of the transitions in the different spectra and normalized them to the intensity of the respective most pronounced transition (see Fig.~\ref{Fig:RelIntensities}). In the following we compared the relative intensities of the different spectra under the assumption that the relative oscillator strengths of the transitions in the pure film spectra do not change upon mixing PFP and PEN if there were no intermolecular coupling. As can be seen from Fig.~\ref{Fig:RelIntensities} there is a change in the relative intensities observable which is a hint for intermolecular coupling in the mixed films, an observation already supported by the existence of the new transition at $E_c=1.6$~eV.
%
%
%
\subsection{Influence of \textit{nonlinear} mixing effects}

\label{Chp:NonlinearModels}
\begin{figure}[h!]
\includegraphics[width=8cm]{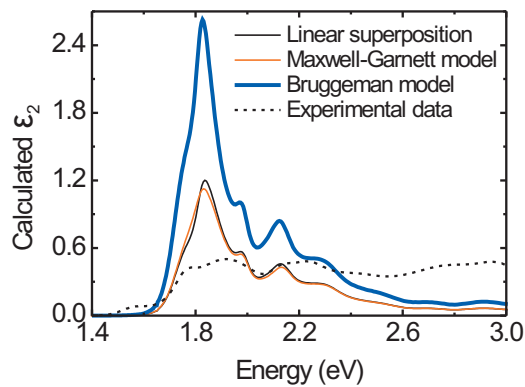}
\caption{(Color online) Comparison of the imaginary part of the dielectric function (in-plane 
component) calculated using different mixing models (linear superposition (thin solid line), Maxwell-Garnett (thin solid line) and Bruggeman (thick solid line)) with the experimental data (dotted line) exemplarily for a 1:1 mixture.}
\label{Fig:Calculatede2}
\end{figure}

If two materials intermix there are different possibilities to describe the effective dielectric function of the mixture without including intermolecular coupling effects. The simplest approach is a linear superposition of the dielectric functions of the single compounds, compare Sec.~\ref{Subsec:Decomp}. In the following we test two models for nonlinear mixing, describing the effective dielectric function of the mixture with an effective medium approximation. The Maxwell-Garnett model~\cite{Garnett_1904_PTRSL_203_385, WVase32_Manual} describes the situation in which inclusions of a material $B$ (with the dielectric function $\tilde{\epsilon}_B$ and volume fraction $f_B$) exist in a host material $A$ (with the dielectric function $\tilde{\epsilon}_A$). In this case the effective dielectric function $\tilde{\epsilon}$ can be calculated as
\begin{equation} 
\frac{\tilde{\epsilon}-\tilde{\epsilon}_A}{\tilde{\epsilon}+ 2\tilde{\epsilon}_A}=f_B\frac{\tilde{\epsilon}_B-\tilde{\epsilon}_A}{\tilde{\epsilon}_B+2\tilde{\epsilon}_A}
\label{Eq:MaxwellGarnett}
\end{equation}
The Bruggeman-Modell~\cite{WVase32_Manual, Bruggeman_1935_AdP_416_636} takes index grading and interface roughness into account and can be used to calculate the effective dielectric function $\tilde{\epsilon}$ of a mixture of the two materials $A$ (with the dielectric function $\tilde{\epsilon}_A$ and volume fraction $f_A$) and $B$ (dielectric function $\tilde{\epsilon}_B$ and volume fraction $f_B$) by solving
\begin{equation}
f_A\frac{\tilde{\epsilon}_A-\tilde{\epsilon}}{\tilde{\epsilon}_A+2\tilde{\epsilon}}+f_B\frac{\tilde{\epsilon}_B-\tilde{\epsilon}}{\tilde{\epsilon}_B+2\tilde{\epsilon}}=0
\label{Eq:Bruggemann}
\end{equation}
Using these models we calculated $\epsilon_2$ for a PFP:PEN 1:1 mixture. The results are shown in Fig.~\ref{Fig:Calculatede2} where it can be clearly seen that these models do not describe the experimental data. Hence we conclude that besides the new transition at $E_c=1.6$~eV, further relevant features result from intermolecular coupling.
%
%
%
\subsection{Influence of scattering}
\begin{figure}
\includegraphics[width=9cm]{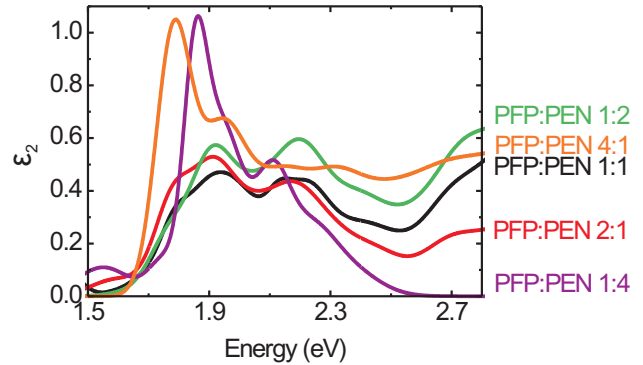}
\caption{(Color online) Imaginary part of the dielectric function (in-plane 
component) of the films with different mixing ratios measured with DRS. The comparison with the VASE-data shows only small differences, see Fig.~\ref{Fig:PFPxPENyinplane}, probably due to photo-oxidation.}
\label{Fig:PFPxPENyDRS}
\end{figure}
In order to investigate the possible influence of scattering from not perfectly smooth surfaces on the spectra of the mixed films in particular for spectral regions above 2.4~eV, we compared spectra obtained with DRS and VASE. While VASE measures the change in the polarization state of polarized light reflected from the sample it is differentely affected by rough surfaces or inhomogeneities in the film than DRS which is sensitive to the absolute intensity. VASE is rather influenced by depolarization effects, whereas scattering decreases the reflected intensity measured with the DRS setup and, therefore, may cause artificial absorption features without pronounced structure in the spectra.
Figure~\ref{Fig:PFPxPENyDRS} shows the spectra of the different mixing ratios of PFP and PEN measured with DRS. Due to the normal incidence geometry of the DRS setup only the in-plane component of the dielectric tensor is probed. The data obtained with DRS are in good agreement with the results of VASE, although minor differences can be observed. These differences could partly be caused by photo-oxidation~\cite{Kytka_2007_APL_90_131911}, but may also be due to practical limitations of the spectral range of the light source used in the DRS-experiment. Comparing both optical techniques, we obtain similar positions of the transitions, comparable line shapes, and an analogous behavior with changing relative amount of PEN and PFP. Independently of the measurement technique, we also observe transitions above 2.4 eV, leading us to the conclusion that these features cannot solely be due to scattering effects.
%
%
%
\section{Out-of-plane component of the absorption spectra}

\begin{figure}
\includegraphics[width=9cm]{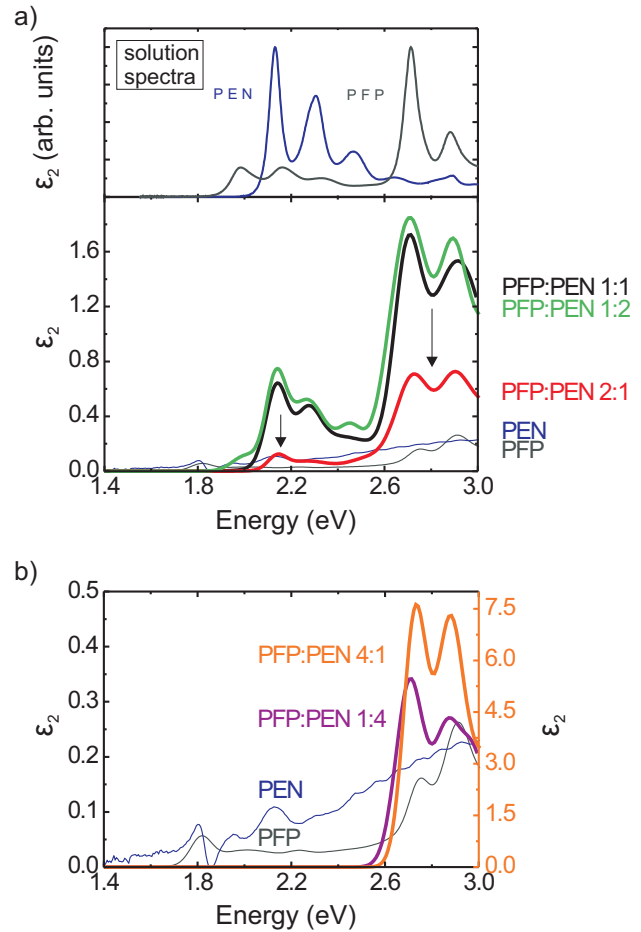}
\caption{(Color online) Imaginary part of the dielectric function (out-of-plane component) of the films for different mixing ratios: (a) PFP:PEN 1:1, 2:1 and 1:2 (thick solid lines). The arrows indicate an increasing amount of PFP. The upper graph shows solution spectra of pure PFP and pure PEN in arbitrary units~\cite{Hinderhofer_2007_J.C.P._127_194705}. (b) PFP:PEN 1:4 (thick solid line, left y-axis) and 4:1 (thick solid line, right y-axis). Note that the absolute amplitudes of the transitions depend on the film thickness $d$ used in the analysis, while $d$ was calibrated using X-ray reflectivity. The uncertainty of the absolute intensity of $\epsilon_2$ is at maximum $\sim 50$\%. The relative error in intensity is below 10\%. For comparison the pure film spectra of PFP and PEN (thin solid lines, left y-axis) similar to Ref.~\onlinecite{Hinderhofer_2007_J.C.P._127_194705} are also shown.}\label{Fig:Anisotropy}
\end{figure}

So far we have only considered the \textit{in-plane} component of the dielectric tensor. Performing a multi-sample analysis~\cite{Heinemeyer_2008_PRB_78_085210} with VASE at different angles we determined also the out-of-plane component. The results are shown in Fig.~\ref{Fig:Anisotropy} and reveal pronounced influence of the intermolecular interactions on the spectra. The features at 2.7 and 2.9~eV observable for all mixing ratios can be related to PFP, since pure PFP exhibits an absorption feature in this energy range with comparable shape but less intense, as Fig.~\ref{Fig:Anisotropy}(a) and (b) shows, which is especially surprising in the case of the mixture with PEN as the dominating compound. As the out-of-plane component is much more affected by errors in the film thickness than the in-plane component, the developement of the absolute intensities could be due to increased disorder in the film and will not be discussed in detail. Even more surprising than the transitions at 2.7 and 2.9~eV are the features in the energy range from 2.14 to 2.44~eV that can only be observed for the mixing ratios 1:1, 2:1 and 1:2 (see Fig.~\ref{Fig:Anisotropy}(a)), but which is missing for the 4:1 and 1:4 mixtures (Fig.~\ref{Fig:Anisotropy}(b)). Since these transitions depend more strongly on the relative amount of PEN and PFP than any of the features in the \textit{in-plane} component does, we speculate that these transitions are caused by intermolecuar interactions between PEN and PFP. Comparing the out-of-plane spectra with solution spectra of the single component films (see Ref.~\onlinecite{Hinderhofer_2007_J.C.P._127_194705}), remarkable similarities can be observed. As the solution spectra show all components of the dielectric tensor at once, the observed similiarities could give a hint for structural changes and a possible reorientation of molecules in the mixed films~\cite{Salzmann_2008_L_24_7294, Hinderhofer_submitted}. Since the shapes of the out-of-plane spectra of mixed films differ so strongly from the ones of pure films, a decomposition into single-film subbands does not provide easily accessible information.

%
\section{Conclusion}

In conclusion, using optical spectroscopy we observed evidence for pronounced intermolecular coupling in mixed films of PEN and PFP with various mixing ratios. The effects of the coupling on the absorption spectra include the appearance of new transitions in both components of the dielectric tensor as well as small blueshifts of the whole spectra presumably arising from a change in the polarisability of the intermolecular environment. In mixed films an increased broadening of the peaks may be caused by inhomogeneities in the film. With a decomposition into single component subbands we performed a first step in the data analysis which enabled us to demonstrate the existence of pronounced \textit{new} transitions in the in-plane component of the mixed films with mixing ratios PFP:PEN 1:1, 2:1 and 1:2, including in particular a new transition at $E_c=1.6$~eV, presumably related to charge transfer. While it is possible to describe the spectral shape of the mixture with PEN dominating approximately by Gaussian oscillators originating from pure PEN, for the mixture with a large amount of PFP this works only in the energy range between 1.7 and 2.2~eV. Besides analyzing the data using a linear mixing model, we tested two models which include \textit{nonlinear} mixing effects.

With the present optical investigations, we have demonstrated that intermolecular interactions have a major impact on the spectra of mixed films containing donor-acceptor interfaces involving molecules of similar shapes.

%
%
%
\begin{acknowledgments}
We thank C. Raisch, University of T\"ubingen, for XPS measurements. Financial support of the DAAD and the DFG is gratefully acknowledged. 
\end{acknowledgments}
%
%

%
\end{document}